\documentclass[a4paper]{nature}
\usepackage[utf8]{inputenc}
\usepackage[english]{babel}
\usepackage[]{graphicx}
\usepackage{amsmath}
\usepackage{amssymb}
\usepackage{mathrsfs}
\usepackage{textcomp}
\usepackage{tabularx}
\usepackage{color}
\usepackage{array}
\usepackage{float}

\definecolor{darkblue}{rgb}{0.1,0.2,0.6}
\definecolor{darkred}{rgb}{0.8,0.1,0.2}
\usepackage[colorlinks,citecolor=darkblue,linkcolor=darkred,urlcolor=darkblue]{hyperref}

\makeatletter
\let\saved@includegraphics\includegraphics
\AtBeginDocument{\let\includegraphics\saved@includegraphics}
\renewenvironment*{figure}{\@float{figure}}{\end@float}
\makeatother

\title{A Double Quantum Dot Spin Valve}

\author{Arunav Bordoloi$^{1,^{*}}$, Valentina Zannier$^{2}$, Lucia Sorba$^{2}$, Christian Sch\"onenberger$^{1,3}$ \& Andreas Baumgartner$^{1,3,^{*}}$}

\begin{document}







\maketitle

\begin{affiliations}
	\item Department of Physics, University of Basel, Klingelbergstrasse 82, CH-4056 Basel, Switzerland
	\item NEST, Istituto Nanoscienze-CNR and Scuola Normale Superiore, Piazza San Silvestro 12, I-56127 Pisa, Italy
	\item Swiss Nanoscience Institute, University of Basel, Klingelbergstrasse 82, CH-4056, Basel, Switzerland
	\item[$^{*}$] Corresponding authors: arunav.bordoloi@unibas.ch, andreas.baumgartner@unibas.ch
\end{affiliations}

\begin{abstract}

A most fundamental and longstanding goal in spintronics is to electrically tune highly efficient spin injectors and detectors, preferably compatible with nanoscale electronics. Here, we demonstrate all these points using semiconductor quantum dots (QDs), individually spin-polarized by ferromagnetic split-gates (FSGs). As a proof of principle, we fabricated a double QD spin valve consisting of two weakly coupled semiconducting QDs in an InAs nanowire (NW), each with independent FSGs that can be magnetized in parallel or anti-parallel. In tunneling magnetoresistance (TMR) experiments at zero external magnetic field, we find a strongly reduced spin valve conductance for the two anti-parallel configurations, with a single QD polarization of $\sim 27\%$. The TMR can be significantly improved by a small external field and optimized gate voltages, which results in a continuously electrically tunable TMR between $+80\%$ and $-90\%$. A simple model quantitatively reproduces all our findings, suggesting a gate tunable QD polarization of $\pm 80\%$. Such versatile spin-polarized QDs are suitable for various applications, for example in spin projection and correlation experiments in a large variety of nanoelectronics experiments.

\end{abstract}

\maketitle


Spin injection and detection are two of the most fundamental processes in semiconductor spintronics,\cite{Das2004,Dietl2014,Fert2017,Baltz2018,Gibertini2019} for example to exploit the electron spin for information storage, logic and sensing,\cite{Datta1990,Dery2007,Parkin2015} or to determine and control spin states in quantum physics.\cite{Jeon2018,Han2019,Nakajima2019} Significant efforts are dedicated to improve the efficiencies of these processes in a variety of material platforms and physical phenomena.\cite{Sahoo2005,Breton2011,Varaprasad2012,Dankert2017,Spaldin2019,Yang2019} Most of these concepts rely on electrical contacts to ferromagnetic reservoirs,\cite{Das2004} or on magnetic tunnel barriers,\cite{Jonker2007} with significant obstacles\cite{Awschalom2007} like a low polarization ($20-40 \%$),\cite{Meservey1994} the magneto-Coulomb effect,\cite{Molen2006,Bernand-Mantel2009} the conductivity mismatch at the metallic ferromagnet-semiconductor interface\cite{Rashba2000} and uncontrolled stray field effects.\cite{Baltz2018} All these effects are particularly challenging in sub-micrometer scaled electronic devices.

\begin{figure}[h]
	\centering
	\includegraphics[width=0.75\columnwidth]{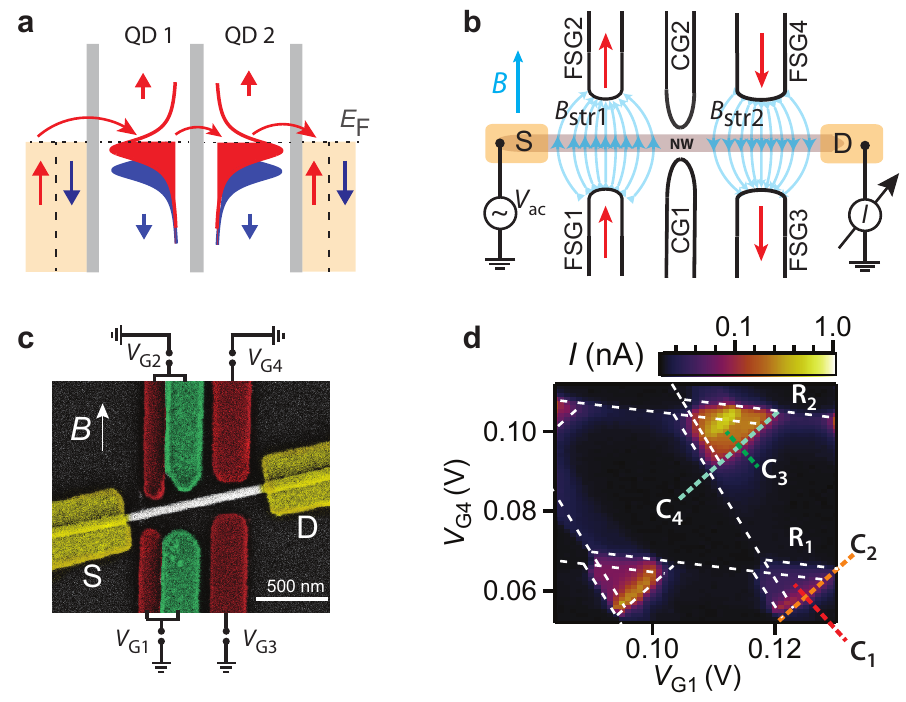}
	\caption{\textbf{Double quantum dot spin valve concept and device.} \textbf{a,b,} Energy diagram (\textbf{a}) and schematic (\textbf{b}) of a DQD spin valve. \textbf{c,} False color SEM image of the investigated InAs NW device. \textbf{d,} Current $I$ as a function of gate voltages $V_{\rm G1}$ and $V_{\rm G4}$, at $V_{\textrm{SD}} =$ +1 mV, showing bias triangles characteristic for weakly coupled DQDs.}
	\label{fig:Fig_1}
\end{figure}

Here we provide an alternative route for spin injection and detection in semiconductor devices using quantum dots (QDs) {\it without} ferromagnetic contacts. As illustrated in figure~\ref{fig:Fig_1}a, the spin degeneracy of a QD state can be lifted by a magnetic field, resulting in a spin polarization at the Fermi energy $E_{\rm F}$ of
\begin{equation}
\label{equ:DefPol}
P = \frac{D_{\uparrow} (E_{\rm F}) - D_{\downarrow} (E_{\rm F})}{D_{\uparrow} (E_{\rm F}) + D_{\downarrow} (E_{\rm F})}, 
\end{equation} 
with $D_{\sigma}$ the QD transmission density of states (t-DoS) for spin state $\sigma  \in \{\uparrow,\downarrow\}$ at $E_{\rm F}$. This spin-dependent transmission directly results in a spin-polarized current through the QD. In practice, a single QD can be spin polarized individually by placing it in the narrow gap in a long strip of a ferromagnetic material, which we term ferromagnetic split-gate (FSG). The FSG generates a stray field $B_{\rm str}$ at the QD position in the direction given by its magnetization, either parallel or antiparallel to its long axis\cite{Fabian2016} and can also be used for electrical gating. The FSG magnetization, and with it $B_{\rm str}$, can be inverted at a characteristic external  switching field $B_{\rm sw}$, determined by the FSG width in the device design.\cite{Aurich2010,Samm2014}

To demonstrate spin injection and detection, we combine two QD-FSG elements in series in a double QD-spin valve (DQD-SV), in which one element acts as spin-injector (polarizer) and the other as spin detector (analyzer). This concept is illustrated in figure~1a: electrons in state $\sigma$ from the {\it unpolarized} electrical contacts tunnel sequentially through the two QDs with a probability that depends on the FSG states of {\it both} QDs, to first order resulting in the respective current $I_{\sigma}\propto D^{(1)}_{\sigma}D^{(2)}_{\sigma}$. Following typical tunneling magnetoresistance (TMR) experiments,\cite{Das2004} we show that in such nano structures both mutually parallel (p) and both anti-parallel (ap) magnetization states of the two FSGs can be accessed at zero external magnetic field, $B=0$, and reoriented by cycling $B$. The individual QD polarizations and TMR signals can be continuously electrically tuned up to values close to the theoretical limits. In contrast to previously employed very large polarizing external magnetic fields,\cite{Folk2003,Potok2002,Hanson2004} the stray and external magnetic fields required for such optimizations are small enough and decay over short enough length scales, to be compatible with various spin injection and detection experiments, for example with superconducting components in Cooper pair splitters\cite{Hofstetter2009,Fueloep2015} for electron spin correlation measurements,\cite{Klobus2014} or to demonstrate equal spin Andreev reflection\cite{He2014} at Majorana type superconducting bound states.\cite{Mourik2012,Deng2016,Zhang2018}


A schematic of a DQD-SV and a scanning electron microscopy (SEM) image of the investigated InAs nanowire (NW) device are shown in figures~\ref{fig:Fig_1}b and ~\ref{fig:Fig_1}c, respectively. The FSGs are long Permalloy (Py) strips fabricated by electron beam lithography with a narrow gap at the NW position, forming the split-gate geometry. The strip widths are $120\,$nm and $230\,$nm, respectively, determining the corresponding switching and stray fields, which can be extracted from independent experiments as demonstrated in Supplementary Information $\textrm{S1}$ and $\textrm{S2}$. The electrical contacts at the NW ends are made of titanium/gold with a split central gate (CG) to electrically form the two QDs fabricated in the same step. One part of the narrower FSG and the CG gate are electrically connected accidentally and are tuned in unison, which we refer to as "gate 1" (G1) and "gate 2" (G2), while the other FSGs are labelled individually (see figure~\ref{fig:Fig_1}c). The DC current $I$ resulting from a bias voltage $V_{\rm SD}$ and the differential conductance $G = \textrm{d}I/\textrm{d}V_{\textrm{SD}}$,  were measured simultaneously using standard DC and lock-in techniques ($V_{\rm ac} = 10\,\mu$V), at a base temperature of $\sim 50\,$mK.


In figure~\ref{fig:Fig_1}d, we plot $I$ flowing through the DQD-SV at $V_{\rm SD}=1\,$mV,  as a function of $V_{\textrm{G1}}$ and $V_{\textrm{G4}}$. This map shows several bias triangles characteristic for a weakly coupled DQD. These triangles originate from one resonance of each QD aligning in energy within the bias transport window.\cite{Wiel2002} This allows us to independently extract most of the single QD parameters used for modelling later, e.g. the lever arms of each gate to each QD (see Supplementary $\textrm{S4}$). We now discuss various types of TMR experiments for two resonances, in figures~\ref{fig:Fig_2} and ~\ref{fig:Fig_3}, respectively, while data for a third resonance are discussed in Supplementary Information $\textrm{S7}$.

We first demonstrate the principle of a TMR experiment and show that all FSG magnetization states can be accessed at $B=0$. Figure~2a shows a high resolution bias triangle of a resonance (not shown in figure~1d) at $V_{\textrm{SD}}=500\,\mu$V. Our typical TMR experiment consists of first choosing a specific trace for the two gate voltages, here by sweeping $V_{\rm G1}$ and keeping $V_{\rm G4}$ constant, as indicated by the red arrow, such that no excited states are involved in the transport process. We then measure $I$ as a function of $V_{\rm G1}$ at a series of external magnetic fields, $B$, applied in parallel to the FSG axes, which results in relatively abrupt switchings of the FSG magnetizations (details in Supplementary Information $\textrm{S2}$). Such a map for the trace in figure~2a is shown in figure~2b for decreasing and increasing magnetic fields, as indicated by the blue and red arrows, respectively, each starting at fields much higher ($+0.5\,$T), or lower ($-0.5\,$T) than shown, to ensure the formation 

\begin{figure}[H]
	\centering
	\includegraphics[width=0.75\columnwidth]{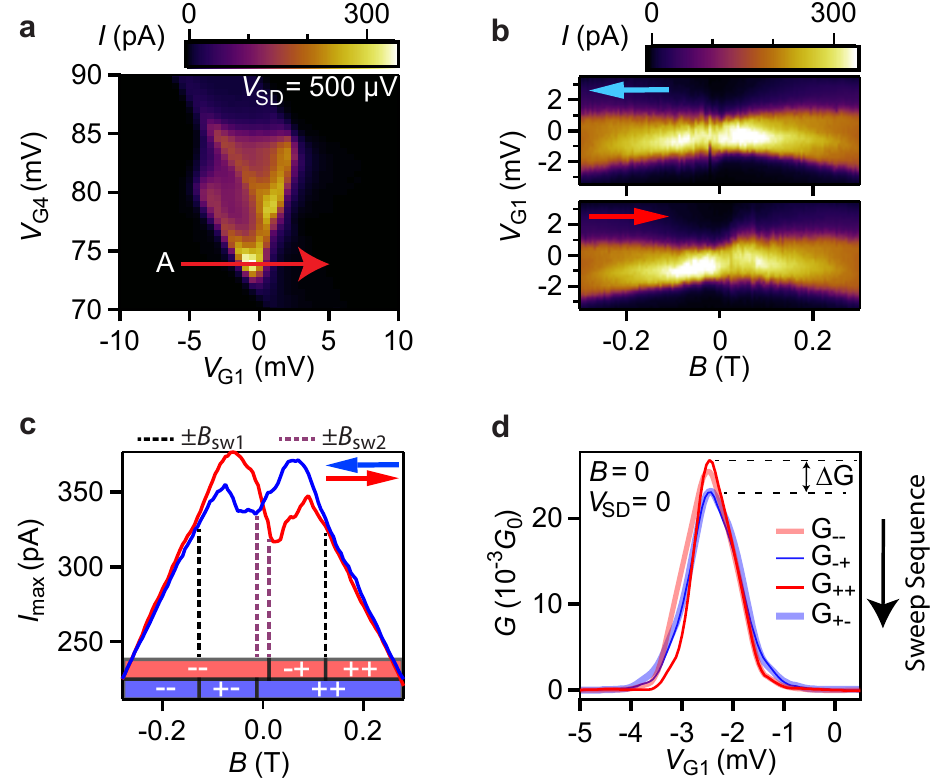}
	\caption{\textbf{FSG magnetization states and TMR at $\mathbf{ \textit{B} = 0}$.} \textbf{a,} Bias triangles at $V_{\textrm{SD}} = 500\,\mu$V. The red arrow specifies the cross-section $\rm A$ investigated in \textbf{b}. \textbf{b,} Up (red arrow) and down sweep (blue arrow) maps of $I$ as a function of $B$ and $V_{\textrm{G1}}$, measured along cross section $\rm A$ in figure~2a. \textbf{c,} Current maximum $I_{\textrm{max}}$ vs $B$, extracted from figure~2b, for the up (red) and down (blue) sweep. The magnetization configurations are indicated by $i,j \in \{+,-\}$ \textbf{d,} $G$ as a function of $V_{\textrm{G1}}$ for all four magnetization states at $B=0$ and $V_{\textrm{SD}}=0$, showing a supression $\Delta G$ for the anti-parallel states relative to the parallel magnetization configurations. The arrow inicates the sequence of the experiments, discussed in detail in Supplementary Information $\rm S3$.}
	\label{fig:Fig_2}
\end{figure}

\noindent of only a single magnetic domain along the FSG axes. These maps show a clear hysteresis with a strong dependence on $B$ and the sweep direction. To demonstrate this more explicitly, we extract the position, width (both discussed in Supplementary Information $\textrm{S5}$) and the maximum current $I_{\rm max}$ at each $B$ value. $I_{\rm max}$ extracted from figure~2b is plotted in figure~2c for decreasing (blue) and increasing $B$ (red).

In the up-sweep, $I_{\rm max}$ first increases roughly linearly with increasing $B$, followed by a maximum at $B \approx -55\,$mT and a decrease around $B=0$. At small positive $B$, $I_{\rm max}$ becomes flatter, followed by a small maximum at $B\approx 85\,$mT, and a roughly linear decrease towards more positive $B$. The down-sweep can be described similarly as the up-sweep, but mirrored at $B\approx 0$ leading to a clear hysteresis. This hysteresis can be understood qualitatively by considering a smooth non-monotonous MR of the DQD that changes abruptly with the reorientation of the FSG magnetizations. In the up-sweep, at $B>B_{\rm sw2}\approx5\,$mT the wider FSG is reoriented parallel to the now positive $B$, and the two FSG magnetizations become anti-parallel (ap). The FSGs become magnetized in parallel again for $B>B_{\rm sw1}\approx 140\,$mT, when the narrower FSG is also inverted (details are given in Supplementary Information $\textrm{S2}$). These configurations are shown schematically at the bottom of figure~2c for the down (blue) and the up-sweep (red).

As a first quantitative measure for the TMR effect, we use the maximum current values at $B = -55\,$mT, using the maximum value of $I_{\rm max}$ in the p state, and the value in the opposite sweep direction at the same field in the ap state. We define TMR as
\begin{equation}
\label{equ:TMRmax}
\textrm{TMR} =\frac{I_{\rm p} - I_{\rm ap}}{I_{\rm p}+I_{\rm ap}} 
\end{equation}
which results in $\textrm{TMR} \approx 6\%$ at $V_{\rm SD} = 500\,\mu$V and $B = -55\,$mT.


To explicitly demonstrate that all four magnetization states (two p and two ap) are accessible at $B=0$, we measure the differential conductance $G$ at $V_{\textrm{SD}} = 0$ as a function of $V_{\textrm{G1}}$ for each FSG magnetization state. The direction of the stray fields $B_{\textrm{str1}}$ and $B_{\textrm{str2}}$ can be reversed individually by sweeping $B$ beyond the characteristic FSG switching fields. For example, we sweep to $B = -500\,$mT and back to $B = 0$ to obtain the $(-,-)$ state, followed by sweeping to $B = +40\,$mT and back to $B = 0$ to obtain the $(-,+)$ state, see Supplementary $\textrm{S3}$ for more details. We note that in the used sequence, p is followed by ap and vice versa. The gate sweeps for the four magnetization states at $B=0$ are plotted in figure~2d. All curves show a maximum at the same gate voltage, which corresponds to a weakly spin split energy level of each QD ($\Gamma>g \mu_{B}B$) being aligned with the Fermi energy. The conductance is gradually reduced to zero if the QD levels are detuned by $V_{\textrm{G1}}$. We find very similar maximum conductances for the same relative magnetization orientations and a clear suppression in $G$ for both ap states with respect to the two p states, yielding $\textrm{TMR}=\frac{\Delta G}{G_{\rm P}+G_{\rm AP}}\approx 7\%$, similar to the value obtained at a larger bias and a small finite $B$.

The DQD-SV experiment can be reproduced quantitatively using a very simple model, which also allows us to estimate the QD polarizations: we assume that the current is given by elastic tunneling in two independent spin channels,\cite{Julliere1975} which yields for a constant weak inter-dot coupling $T_{12}$ and the magnetization orientations $i,j \in \{+,-\}$ along the FSG axes, 
\begin{equation}
\label{equ:I_TMR}
\begin{split}
I^{(ij)} = I^{(ij)}_{\uparrow} + I^{(ij)}_{\downarrow} = &\frac{e}{h} \sum_{\sigma} \int_{-\infty}^{\infty}  T_{12} D_{1 \sigma}^{(i)} (E) D_{2 \sigma}^{(j)} (E) \\
&[f(E-\mu_{\rm S})-f(E-\mu_{\rm D})]dE,
\end{split}
\end{equation}
where $D_{\beta \sigma} (E)$ denotes the spin dependent t-DoS in dot $\beta \in \{1,2\}$ and $\sigma \in \{\uparrow, \downarrow\}$ the spin orientation; $f(E)=1/(1+ e^{E/(k_{B}T)})$ is the Fermi-Dirac distribution function and $\mu_{\rm S,D}$ the electrochemical potential in the source and drain contacts, respectively. To start with, we assume a small bias (linear regime) to obtain the conductance, as in the experiments. Since the Zeeman shift is opposite, but of the same magnitude for opposite spins, the t-DoS of each QD obeys the identity $D_{ \sigma}^{-} (-B,E_{\rm F}) = D_{ -\sigma}^{+} (+B,E_{\rm F})$ due to time-reversal symmetry. At $B=0$, this reduces to $D_{ \sigma}^{-} (E_{\rm F}) = D_{ -\sigma}^{+} (E_{\rm F})$, which yields, using the definition of the QD polarizations in equation~(1),
\begin{equation}
\label{equ:Polarization}
\textrm{TMR} = \frac{I_{\textrm{p}} - I_{\textrm{ap}}}{I_{\textrm{p}} + I_{\textrm{ap}}} = P_{1} P_{2} \approx P^{2}. 
\end{equation}         
In the last step we assume that both QD polarizations are identical, which results in $P \approx 27\%$ on resonance at $B = 0$. We stress that this expression for the TMR signal only holds at $B=0$ because of the non-constant QD t-DoS, in contrast to devices with ferromagnetic contacts, for which it holds also at finite external fields, limited only by the correlation energy of the band structure.


The non-constant t-DoS of the QDs allows us to go beyond the standard experiments, enabling us to optimize and tune the TMR signals magnetically as well as electrically. To demonstrate this, we investigate cross section $\rm C_{1}$ pointed out in figure~1d, for which we again plot $I$ as a function of $B$ and $V_{\textrm{G1}}$ at $V_{\textrm{SD}} = 10 \mu\textrm{V}$. Figure~3a shows the up and down sweeps, which again show a clear hysteresis, prominently visible in figure~3b, where we plot $I_{\textrm{max}}$ as a function of $B$ for the up and down sweeps (width and position are discussed in Supplementary Information $\textrm{S5}$). These curves show qualitatively similar characteristics as discussed for figure~2c. From the current maximum, we find a TMR signal of $\sim 29\%$ at $B=0$ and estimate the individual QD spin polarizations as $P \approx 53 \%$ using equation~(4). These values are larger than for the previously discussed resonance, mostly due to a smaller resonance width.

\begin{figure}[H]
	\centering
	\includegraphics[width=0.75\columnwidth]{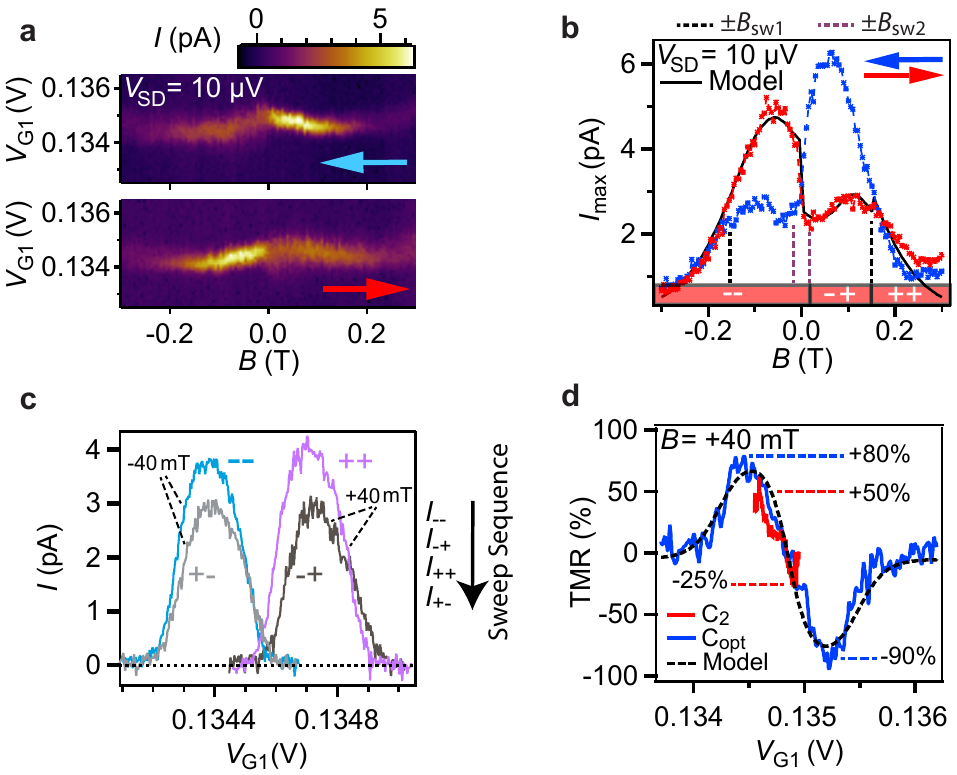}
	\caption{\textbf{Optimized TMR at $\mathbf{\textit{B} = \pm 40\,}$mT.} \textbf{a,} Maps of $I$ as a function of $B$ and $V_{\textrm{G1}}$ for the up (red arrow) and down sweep (blue arrow), for the cross section $\rm C_{1}$ (see figure~1d) at $V_{\textrm{SD}}=10\,\mu$V. \textbf{b,} Maximum current $I_{\textrm{max}}$ as a function of $B$ for the up (red) and down sweep (blue) extracted from figure~3a. \textbf{c,} $I$ along cross section $\rm C_{1}$ (see figure~\ref{fig:Fig_1}d and details in Supplementary Information S6) paramterized by $V_{\textrm{G1}}$ for all four magnetization states, with the $(-,-)$ and $(+,-)$ configurations measured at $B=-40\,$mT, and the $(+,+)$ and $(-,+)$ configurations at $B=+40\,$mT. \textbf{d,} TMR for magnetization states $(+,+)$ and $(-,+)$ at $B = +40\,$mT for cross sections $\rm C_{2}$ (red) and $\rm C_{\rm opt}$ in figure~\ref{fig:Fig_4}a (blue). The black dashed line shows the TMR extracted from the model for cross section $\rm C_{\rm opt}$ (shown in figure~\ref{fig:Fig_4}a), with the parameters obtained from fits to the data in figure~\ref{fig:Fig_3}b.}
	\label{fig:Fig_3}
\end{figure}

We now exploit the non-constant t-DOS to optmize the TMR signal. First, we apply a small homogenous external field of $\pm 40\,$mT, which is small enough to still access all four FSG magnetization states ($B < B_{\rm sw1}$) and compatible with a wide variety of applications, for example with many superconducting circuit elements. We measure $I$ along cross section $\rm C_{2}$ indicated in figure~1d, which is chosen on the resonance maximum along the base of the bias triangle (see Supplementary Information $\textrm{S6}$) so that a shift in the resonance energies is negligible. 

Figure~3c shows the four $I (V_{\textrm{G1}})$ curves along $\rm C_{2}$ for the four FSG magnetization states $(i,j)$ ($V_{\rm G4}$ is the same for each chosen $B$). The curve for the parallel $(-,-)$ [blue] and the anti-parallel configuration $(+,-)$ [grey] were measured at $B=-40\,$mT, while the ones for $(+,+)$ [purple] and (-,+) [black] were measured at $B=+40\,$ mT (see Supplemntary Information S3 for sweep sequence). We find that the maximum current and lineshape for both anti-parallel configurations are almost identical, while the two parallel ones slightly differ. Most importantly, the anti-parallel curves are reduced in amplitude by $\sim 25\%$ with respect to the parallel ones. We note that for this cross section, the maximum occurs at the same $V_{\rm G1}$ value for both pairs of curves in figure~\ref{fig:Fig_3}c. 

\begin{figure}[H]
	\centering
	\includegraphics[width=0.8\columnwidth]{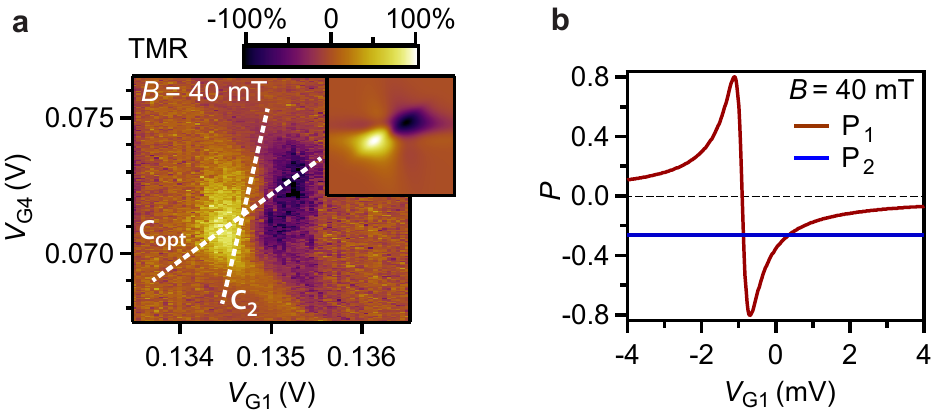}
	\caption{\textbf{From TMR to spin polarization.} \textbf{a,} Measured TMR as a function of $V_{\textrm{G1}}$ and $V_{\textrm{G4}}$ for the magnetization states $(+,+)$ and $(-,+)$ at $B = 40\,$mT. The cross sections $\rm C_{2}$ and $\rm C_{\rm opt}$ are indicated by dashed lines. Inset: TMR from the model calculations with the parameters extracted from figure~\ref{fig:Fig_3}b . \textbf{b,} Spin polarization of QD1 $(P_{1})$ and QD2 $(P_{2})$ as a function of $V_{\textrm{G1}}$ and constant $V_{\textrm{G4}}$ from the model at $B=40\,$mT, showing a large gate tunability of $P_{1}$ from $-0.8$ to $+0.8$.}
	\label{fig:Fig_4}
\end{figure}

For any given $V_{\rm G1}$ and $B$, we now calculate the TMR signal using equation~(\ref{equ:TMRmax}). As an example, this is plotted for the states $(+,+)$ and $(-,+)$ in figure~3d (red curve), which shows that the TMR signal is continuously gate tunable roughly between $+50\%$ and $-25\%$. This TMR signal can be improved significantly by exploiting the small, field-induced shifts in the resonance positions. To achieve this, we plot $\textrm{TMR} = (I_{++} - I_{-+})/(I_{++} + I_{-+})$ at $B = 40\,$mT as a function of $V_{\rm G1}$ and $V_{\rm G4}$ in figure~\ref{fig:Fig_4}a and find the optimal cross section labelled $\rm C_{\rm opt}$. In figure~\ref{fig:Fig_3}d, we plot TMR along $C_{\rm opt}$ which shows a continously gate tunable TMR with a well separated pronounced maximum and minimum TMR of $+80\%$ and $-90\%$, respectively. These values are significantly larger than in most other systems.

We expect that the QD polarizations are also gate tunable to large values, but since an external field is applied, the above symmetry argument cannot be used for a simple estimate. We therefore resort to numerically evaluating the model introduced above. To do so, we define the total magnetic fields $B_{\textrm{tot}}^{(\beta)} = B + B_{\textrm{str}}^{(\beta)}$ at the two QD positions $\beta\in \{1,2\}$, and use as the energy-dependent t-DoS of the QDs at energy $E$ the Lorentzian $\mathcal{L} (E-E_{\beta \sigma}) = (\Gamma_{\beta}/2)^2/[(E - E_{\beta \sigma})^2 + (\Gamma_{\beta}/2)^2]$, centered at
\begin{equation}
\label{equ:DOS_QD}
E_{\beta \sigma} = E_{\beta}^{(0)} - e \alpha_{\beta} V_{g \beta}  + \frac{1}{2} \sigma g_{\beta} \mu_{B} B_{\textrm{tot}}^{(\beta)},
\end{equation}        
with $E_{\beta}^{(0)}$ an energy offset for states in dot $\beta$ at zero gate voltages, $g_{\beta}$ the corresponding electron g-factors and $\Gamma_{1}$ and $\Gamma_{2}$ the broadening parameters. The lever arms $\alpha_{\beta}$ are extracted independently from the bias triangles\footnote[2]{we also include cross lever arms in the model without stating this explicitly for simplicity} and $V_{{\rm g} \beta}$ are the applied gate voltages. The total current is then calculated using equation~(\ref{equ:I_TMR}).

This model reproduces very well the experiments using a single set of parameters for a given resonance, all in the typical range found in literature. For example, we obtain $I_{\textrm{max}}$ as a function of $B$, as plotted by the black curve in figure~\ref{fig:Fig_3}b for the up sweep, using $B_{\textrm{str1}} = 61 (\pm 4)\,$ mT, $B_{\textrm{str2}} = 27 (\pm 5)\,$ mT (estimated independently, see Supplementary S1) and the adjustable parameters $g_{1}= 5.6$, $g_{2} = 6.3$, $\Gamma_{1}=25\,\mu$eV and $\Gamma_{2}=15\,\mu$eV, $E_{1}^{(0)} \equiv 0$, $E_{2}^{(0)} = 8.1 (\pm 0.3)\, \mu$eV, and an inter-dot tunnel coupling $T_{12}=0.12$ adjusted to obtain the correct amplitude. The errors given in brackets indicate the range for a parameter that still gives satisfactory model curves. The same parameters also reproduce the TMR results, shown as an inset in figure~\ref{fig:Fig_4}a and the optimized TMR cross section $\textrm{C}_{5}$ shown in figure~\ref{fig:Fig_3}d (black dashed line). The same parameters also reproduce the width (Supplementary Information $\textrm{S5}$) and figure~\ref{fig:Fig_3}c. To reproduce the other investigated resonances, we use slightly different parameters, as summarized in Supplementary Information $\textrm{S7}$.

In the model it is straight forward to extract the spin polarizations, e.g. $P_{1}$ for QD1 as a function of $V_{\textrm{G1}}$ at $B=40\,$mT, which is plotted in figure~\ref{fig:Fig_4}b, with $P_{2}\approx 27\%$ for QD2, being independent of $V_{\rm G1}$. $P_{1}$ can be gate tuned over a large range, with a maximum absolute value of $ P_{1} \approx 80 \%$, and a zero-field value of $\approx 59\%$. This analysis demonstrates that the DQD-SV is a highly tunable spin valve with one QD acting as a gate-tunable spin injector and the other as a detector, such that transport through the DQD can be electrically tuned from predominantly spin down electrons to spin up electrons, depending on the orientation of $B_{\textrm{str}}$ and $B$. The large gate-tunability of the QD spin polarizations originates from the resonance widths being of similar magnitude as the Zeeman splitting, $\Gamma_1+\Gamma_2 \sim g\mu B_{\textrm{tot}}$. Increasing the QD life time in the model by only a factor of two, keeping all other parameters the same, we find even stronger polarizations, up to 91$\%$, thus almost reaching unity. Such sharper line shapes can be obtained with in situ grown InP tunnel barrier\cite{Fuhrer2007,Roddaro2011,Thomas2019}  or by crystal phase engineered barriers in InAs NWs.\cite{Juenger2019,Nilsson2016} In addition, the QD polarization can be enhanced by stronger $\rm B_{str}$, either by reducing the FSG gap, e.g. using smaller diameter NWs, or by using other ferromagnetic materials.

In conclusion, we have demonstrated a DQD spin valve in an InAs NW with ferromagnetic split gates that results in a tunneling magnetoresistance electrically tunable between $+80 \%$ and $-90\%$. Using a simple resonant tunneling model, we extract gate and $B$ field tunable QD spin polarizations up to $\sim \pm 80 \%$, with the possibility of even larger values, up to unity. The small external fields resulting in such large efficiences are compatible with many superconducting contacts in close proximity,\footnote[3]{see supplementary information S1 for an experimental estimate of $B_{\rm str}$ away from the FSGs.} so that the QD-FSG units are ideally suited as spin injectors and detectors in nanoelectronic devices, for example to investigate spin orbit interactions, to perform spin correlation measurements and electronic Bell tests in a Cooper pair splitter,\cite{Klobus2014} or to demonstrate equal spin Andreev reflection at Majorana zero modes.\cite{He2014} In addition, an array of such FSG units could in principle be used to engineer a synthetic and externally controllable spin orbit interaction.\cite{Braunecker2010,Kjaergaard2012}

\bibliography{Reference_DQD}

\section*{Acknowledgements}

This work has received funding from the Swiss National Science Foundation, the Swiss Nanoscience Institute, the Swiss NCCR QSIT, the FlagERA project Topograph, the QuantERA SuperTop project network and the FET Open project AndQC. C.S. has received funding from the European Research Council under the European Union's Horizons 2020 research and innovation programme.
\section*{Author Contributions}

A.Bo fabricated the devices, performed the measurements and analyzed the data. V.Z. and L.S. have grown the nanowires. A.Ba provided the model and helped with the measurements and data analysis. A.Bo and A.Ba wrote the paper. C.S. and A.Ba initiated and supervised the project. All authors discussed the results and contributed to the manuscript.   

\section*{Competing Interests}

The authors declare no competing interests.

\section*{Additional Information}

\noindent All data in the publication are available in numerical form at DOI: \url{https://doi.org/10.5281/zenodo.3557857}.

\section*{METHODS}

\noindent The InAs NWs were grown using 30 nm gold (Au) colloid assisted chemical beam epitaxy\cite{Gomes2015} and have a diameter of 40-45 nm and a length of 2.0-2.3 $\mu$m. The NWs were mechanically transferred from the growth substrate to a heavily p-doped silicon substrate serving as a global backgate (BG), with a 400 nm $\rm SiO_{2}$ insulating top layer. For the electron beam lithography, we employed pre-defined markers and contact pads made of Ti/Au (5 nm/ 45 nm). The central gates and electrical contacts at the NW ends were first made of Ti/Au (5 nm/ 45 nm), while the ferromagnetic split-gates were fabricated in a second step and made of 30 nm thick permalloy (Py), while. Before evaporating the contact material, the native oxide of the NWs is etched with a 1:10 ratio $(\rm NH_{4})_{2}S_{x} : H_{2}O$ solution for 3.5 minutes. The $(\rm NH_{4})_{2}S_{x}$ solution was prepared by mixing 0.96 grams of sulfur powder in 10 ml of ammonium sulfide solution (20$\%$ in $\rm H_{2}O$).

\end{document}


\maketitle

\section{Determination of the stray field in the FSG gap}

\begin{figure}[h]
	\begin{center}
		\includegraphics[width=0.85 \columnwidth]{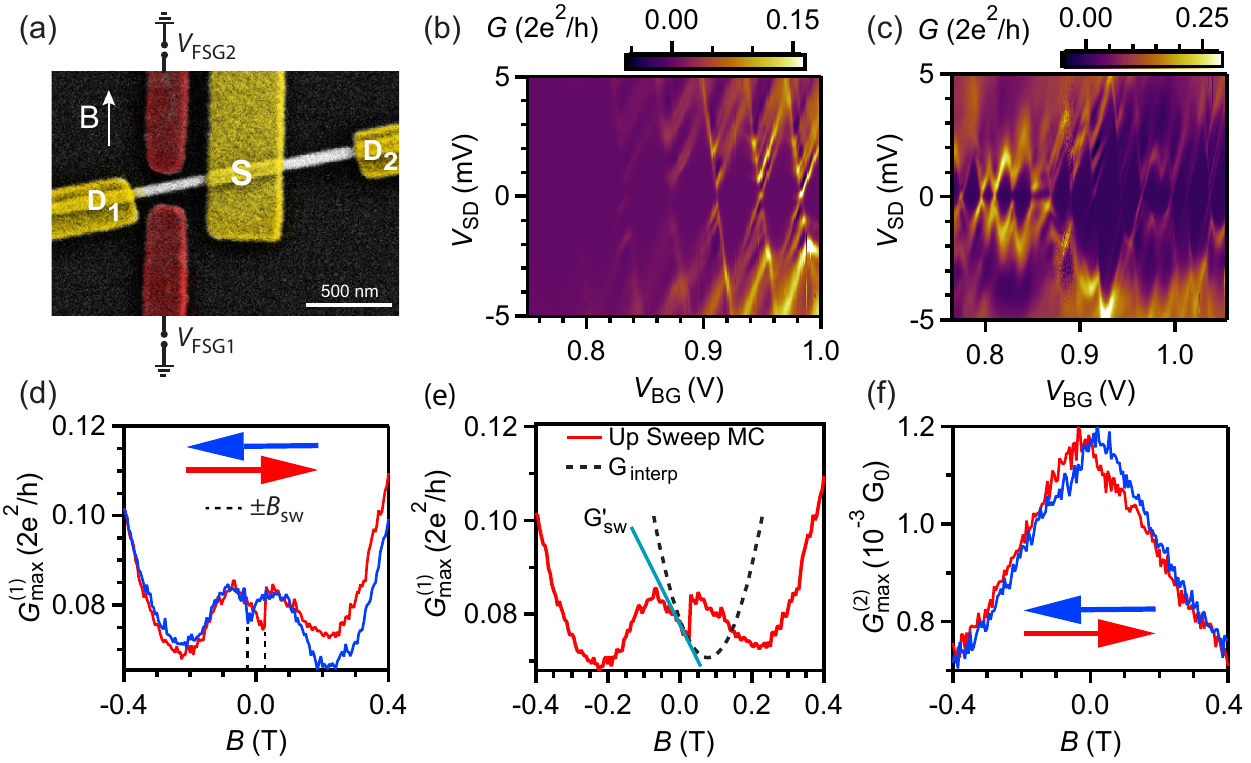}
	\end{center}
	\caption{\textbf{Determination of FSG stray field.} (a) False color SEM image of a 3-terminal NW device. The FSG is placed at a distance of 35 nm from the NW. (b,c) Differential conductance $G$ as a function of dc bias $V_{\textrm{SD}}$ and the back gate voltage $V_{\textrm{BG}}$ showing Coulomb blockade diamonds for the NW segment (b) $S-D_{2}$ and (c) $S-D_{1}$ respectively. (d) Maximum conductance $G_{\textrm{max}}$ as a function of external magnetic field $B$, applied along the FSG long axes, for the up (red) and down (blue) sweep of a QD resonance, with $V_{\rm SD} = 0$, in the NW segment $S-D_{1}$. (e) $G_{\textrm{max}}$ vs $B$ for the up sweep showing the slope ($G'_{\rm sw}$) at $B = B_{\rm sw}$ and the interpolated parabola ($G_{\rm interp}$), resulting in $B_{\textrm{str}} \approx 80$ mT. (f) $G_{\textrm{max}}$ vs $B$ for the up (red) and down (blue) sweep of a QD resonance, at $V_{\rm SD} = 0$, in the NW segment $S-D_{2}$.}	
	\label{bstr}
\end{figure}

As a control experiment to the DQD-SV, we fabricate a device with a single pair of FSG as shown in figure \ref{bstr}a. The device is a 3-terminal InAs NW contacted with Ti/Au normal metal contacts S, $\rm D_{1}$ and $\rm D_{2}$. Adjacent to the NW segment between $S$ and $D_{1}$, we placed a 170 nm wide Permalloy FSG pair in a split gate geometry at a distance of 35 nm from the NW. We apply a dc ($V_{\rm SD}$) and ac bias to the source $(S)$ contact and simultaneously measure the differential conductance $G^{(1,2)} = \frac{dI_{1,2}}{dV}$ at contacts $D_{1}$ and $D_{2}$ using standard lock-in techniques.

Figures \ref{bstr}b and \ref{bstr}c show colorscale plots of $G^{(1)}$ and $G^{(2)}$, respectively as a function of $V_{\textrm{SD}}$ and back gate voltage $V_{\textrm{BG}}$. We observe a regular pattern of Coulomb Blockade (CB) diamonds, suggesting the formation of QDs in both NW segments. We then apply an external magnetic field $B$ along the FSG long axes and measure $G^{(1),(2)}$ as a function of $B$ for the QD resonances in the $S-D_{1}$ and $S-D_{2}$  segment, similar to experiments in the main text. 

The maximum conductance $G^{(1)}_{\textrm{max}}$ over one CB resonance ($V_{\rm BG} = 1.2\,$V) as a function of $B$ is shown in figure \ref{bstr}d. It is clearly hysteretic for the up (red) and down (blue) sweeps and mirror symmetric around $B=0$. We also observe a sharp switching in this magnetoconductance (MC) at $B_{\rm sw} = 25\,$mT for both sweeps, suggesting a reversal in the magnetization of the FSG. We note that we only observe a single switching, suggesting that the two parts of the FSG switch in unison. Furthermore, for both sweeps in the NW segment $S-D_{2}$, we observe a small hysteresis and no switching, as shown in figure~\ref{bstr}f. We note that we do not observe any hysteresis in the magnetoconductance for QD devices without any FSGs nearby (not shown).

For a two terminal device with $B_{\rm str} = 0$, the MC $G_{0}(B)$ is necessarily an even function in $B$

\begin{equation}
\label{equ:bstray}
G_{0}(-B) = G_{0} (+ B) 
\end{equation}     
This relation allows us to determine the stray field of the FSGs assuming $\vec{B}_{\rm str} || \vec{B}$. For a given $G_{0}(B)$, we can define the $B_{\rm str} \ne 0$ curves as $G_{-} (B) = G_{0}(B-B_{\rm str})$ for the up sweep up to $B \leq +B_{\rm sw}$ and $G_{+} (B) = G_{0}(B+B_{\rm str})$ for the down sweep down to $B \geq -B_{\rm sw}$. From equation~\ref{equ:bstray}, we find $G_{+} (B) = G_{-} (-B)$ and $G_{+} (B) = G_{-} ( B + 2B_{\rm str})$.

 We now use a polynomial (or any other suitable function) to interpolate the data of $G_{-} (B)$ up to $+ B_{\rm sw}$ and use the same function to extrapolate to the next extremum, where $\frac{dG_{-}(B)}{dB}|_{B = B_{str}} = 0$ by symmetry, allowing us to directly read off $B_{\rm str}$. For a simple analytical estimate of $B_{\rm str}$, we use the lowest order even polynomial $G_{\rm interp} = a(B - B_{\rm str})^2 + b$, with the curvature $a$ fixed to the maximum observed value in the experiment. Using the slope at $B = B_{\rm sw}$ obtained from the experiment, one finds $G'_{\rm sw} = \frac{dG_{-}}{dB}|_{B = B_{\rm sw}} = 2a(B_{\rm sw} - B_{\rm str})$, assuming $B_{\rm str} > B_{\rm sw}$, which results in $B_{\rm str} = B_{\rm sw} - \frac{G'_{\rm sw}}{2a}$. The parameter $b$ is not used here, but can be obtained by matching $G_{-}$ and $G_{interp}$ at $B = B_{\rm sw}$. The method is illustrated in figure~\ref{bstr}e. A similar analysis also holds for the down sweep $G_{+} (B)$. With this method, we find a lower bound for $B_{\rm str} \approx 80\,$mT. A similar analysis for the $S-D_{2}$ segment results in $B_{\rm str} \leq 5\,$mT, consistent with the much larger distance from the FSG.


For our TMR device in the main text, we extract $B_{\rm str,QD1} = 60\,$mT and $B_{\rm str,QD2} = 25\,$mT from the resonant tunneling model, consistent with the larger distances of the FSGs from the InAs nanowire, with the gates G1 and G4 being 55 nm and 90 nm away from the nanowire respectively.

\section{Determination of $\textrm{B}_{\textrm{switch}}$}

\begin{figure}[H]
	\centering
	\includegraphics[width= 0.7\textwidth]{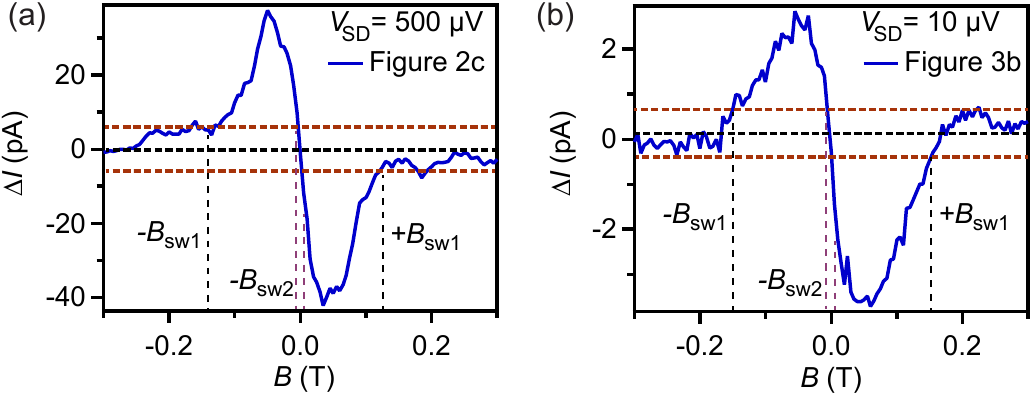}	
	\caption{\textbf{Determination of FSG switching fields.} (a) $\Delta I = I_{\rm up}^{\rm max} - I_{\rm down}^{\rm max}$ as a function of $B$ for the experiment in Figure 2c at $V_{\textrm{SD}} = 500\,\mu$V. (b) $\Delta I$ as a function of $B$ for the experiment in Figure 3b at $V_{\textrm{SD}} = 10\,\mu$V. }
	\label{Bswitch}
\end{figure}

To determine the characteristic switching fields $B_{\rm sw}$ of the FSGs in the DQD-SV, we plot $\Delta I = I_{\rm up}^{\rm max} - I_{\rm down}^{\rm max}$ as a function of $B$ in figure~\ref{Bswitch}a, where $I_{\rm up}^{\rm max}$ and $I_{\rm down}^{\rm max}$ refer to the up and down sweep in figure~2c of the main text, measured at $V_{\rm SD} = 500\,\mu$V. We assign an average zero level of the measured data, shown as black dashed line and define a lower and upper current limit for a significant deviation of $\Delta I$ from the average zero. We use the $B$ values at which the upper horizontal line meets $\Delta I$ as the two switching fields, $B_{\textrm{sw1}} \approxeq 140\,$mT and $B_{\textrm{sw2}} \approxeq 5\,$mT, respectively. We use a similar analysis for the lower horizontal line. We point out that a similar analysis of Figure 3b, i.e. on a different resonance, results in the same switching fields, as shown in figure~\ref{Bswitch}b.

\section{Sweep Sequence}

The determination of the switching fields enables us to define the exact procedure to obtain the four magnetization states at $B = 0$. The measurements in the main text were all done in the following orders:

\begin{enumerate}
\item (-,-): Sweep the external magnetic field to $B = -500\,$mT $<< -B_{\rm sw1}$ in order to form a single magnetic domain along the FSG axis, followed by a sweep back to $B = 0$ to obtain the magnetization state $(-,-)$.
\item  (-,+): Continue sweeping to $B = +40\,$mT $> B_{\rm sw2}$ (but $< B_{\rm sw1}$) followed by a sweep back to $B = 0$ to obtain the magnetization state $(-,+)$.
\item (+,+): Sweep to $B = +500\,$mT $>> B_{\rm sw1}$ to get a single magnetic domain along the $+B$ direction, followed by a sweep back to $B = 0$ to obtain $(+,+)$.
\item (+,-): Continue sweeping to $B = -40\,$mT $< -B_{\rm sw2}$ (but $> -B_{\rm sw1}$) followed by a sweep back to $B = 0$ to obtain $(+,-)$.
\end{enumerate}

Similarly, the field sweep sequences used in the experiments on the four magnetization states at $B = \pm 40\,$mT in the main text are as follow:

\begin{enumerate}
	\item (-,-): Sweep the external magnetic field to $B = -500\,$mT in order to form a single magnetic domain along the FSG axis, followed by a sweep back to $B = -40\,$mT to obtain the magnetization state $(-,-)$ at $B = -40\,$mT.
	\item (-,+): Sweep to $B = +40\,$mT to obtain the magnetization state $(-,+)$ at $B = +40\,$mT.
	\item (+,+): Continue sweeping to $B = +500\,$mT to get a single magnetic domain along the $+B$ direction, followed by a sweep back to $B = +40\,$mT to obtain the $(+,+)$ at $B = +40\,$mT.
	\item (+,-): Continue sweeping to $B = -40\,$mT to obtain $(+,-)$ at $B = -40\,$mT.
\end{enumerate}

\section{DQD Characterization}

\begin{figure}[H]
	\begin{center}
		\includegraphics[width= 0.45 \textwidth]{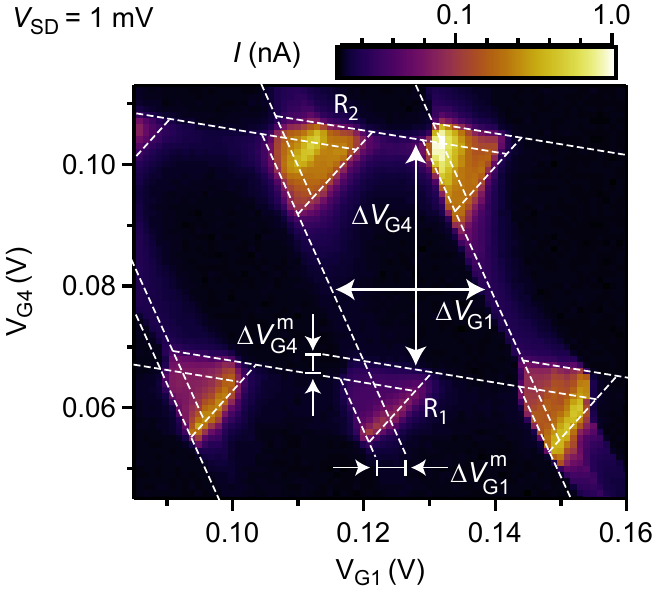}
	\end{center}
	\caption{ \textbf{Characterization of the double quantum dot.} Colorscale plot of the current $I$ as a function of $V_{\textrm{G1}}$ and $V_{\textrm{G4}}$ at $V_{\textrm{SD}} = 1$ mV.}
	\label{DQD_charac}
\end{figure}

Figure \ref{DQD_charac} shows the current $I$ as a function of $V_{\textrm{G1}}$ and $V_{\textrm{G4}}$ at $V_{\textrm{SD}} = 1$ mV to characterize the weakly-coupled serial DQD. The other gates are kept constant at $V_{\textrm{BG}} = -0.25$ V, $V_{\textrm{G2}} = -0.2$ V and $V_{\textrm{G3}} = 0.0$ V. From the honeycomb structure \cite{Wiel2002}, we obtain $\Delta V_{\textrm{G1}} = 25$ mV, $\Delta V_{\textrm{G4}} = 30$ mV, $\Delta V^{\textrm{m}}_{\textrm{G1}} = 3$ mV and $\Delta V^{\textrm{m}}_{\textrm{G4}} = 4$ mV, as shown in figure~\ref{DQD_charac}. The capacitance between the QD and the respective gate is given by: $C_{\rm G} = e/\Delta V_{\rm G}$. We find $C_{\textrm{G1}} = 6.4$ aF and $C_{\textrm{G4}} = 5.94$ aF to the respective QD. The total capacitances of the two QDs are $C_{1} = 64$ aF and $C_{2} = 65.2$ aF, while the mutual capacitance is $C_{\textrm{m}} = 7$ aF. The addition energy of the QDs are $E_{\textrm{add,1}} \approx 2.5$ meV and $E_{\textrm{add,2}} \approx 2.7$ meV, while the level spacings are $\delta E_{\textrm{1}} \approx 0.7$ meV and $\delta E_{\textrm{2}} \approx 0.81$ meV, respectively. The lever arms for both dots are found as: $a_{11} \approx 0.1$, $a_{12} \approx 0.015$, $a_{21} \approx 0.0$ and $a_{22} \approx 0.09$, i.e. the cross lever arms are one order of magnitude smaller than the direct lever arms.

\section{Width and Position of the DQD-SV resonances}

\begin{figure}[H]
	\centering
	\includegraphics[width= 0.6\textwidth]{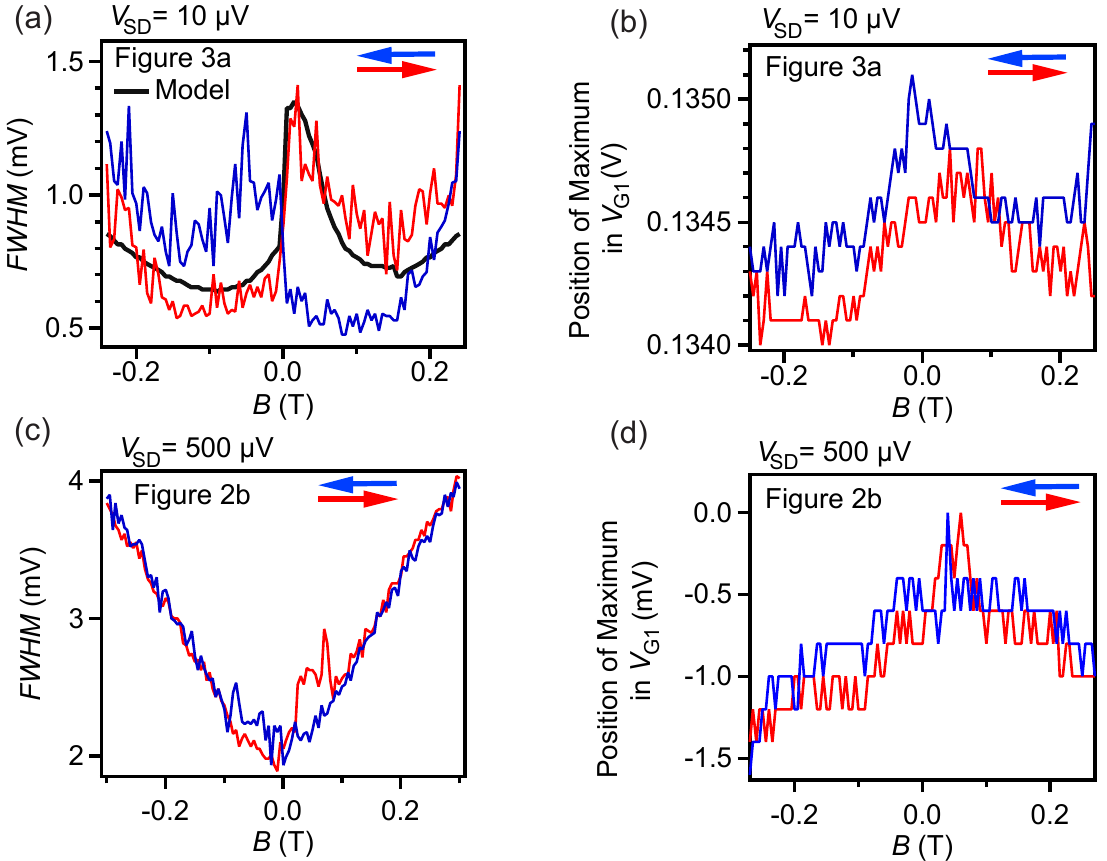}	
	\caption{\textbf{FWHM and position of the resonance maximum.} (a) Full width at half maximum (FWHM) of the DQD-SV resonance as a function of $B$ for the up (red) and down (blue) sweep for the experiments in Figure~3a at $V_{\textrm{SD}} = 10\,\mu$V. The black line shows the FWHM obtained from the resonant tunneling model using the same parameters as in the main text. (b) Position of the current maximum in $V_{\rm G1}$ as a function of $B$ for the up (red) and down (blue) sweep for experiments in Figure~3a at $V_{\textrm{SD}} = 10\,\mu$V. (c) FWHM as a function of $B$ for the up (red) and down (blue) sweep for experiments in Figure~2b at $V_{\textrm{SD}} = 500\,\mu$V. (d) Position of the current maximum in $V_{\rm G1}$ as a function of $B$ for the up (red) and down (blue) sweep for experiments in Figure~2b.}
	\label{width}
\end{figure}




\section{Bias Triangle for the Four Magnetization States}

\begin{figure}[H]
	\centering
	\includegraphics[width= 0.6\textwidth]{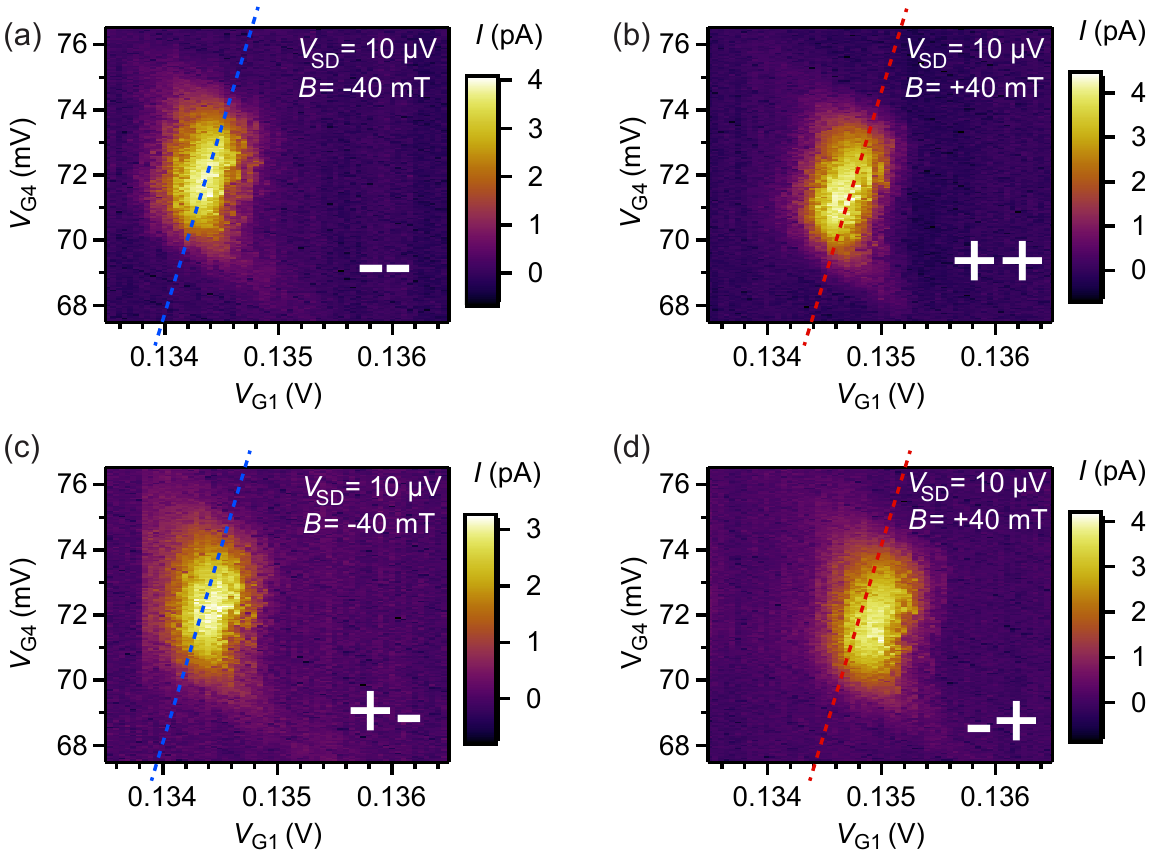}	
	\caption{\textbf{Bias triangles for four magnetization states.} $I$ as a function of $V_{\textrm{G1}}$ and $V_{\textrm{G4}}$ for the magnetization state (a) $(-,-)$ at $B = -40\,$mT (b) $(+,+)$ at $B = +40\,$mT (c) $(+,-)$ at $B = -40\,$mT and (d) $(-,+)$ at $B = +40\,$mT, measured at $V_{\textrm{SD}} = 10 \mu$V. The dashed red and blue lines are guide to the eyes.}
	\label{bias_triangle}
\end{figure}

The current $I$ as a function of $V_{\textrm{G1}}$ and $V_{\textrm{G4}}$ for the four magnetization states at $B = \pm 40\,$mT and $V_{\textrm{SD}} = 10 \mu$V is shown in Figure~\ref{bias_triangle}. The $(-,-)$ and $(+,-)$ magnetization states were measured at $B = -40\,$mT, while the $(+,+)$ and $(-,+)$ magnetization states were measured at $B = +40\,$mT. For the same $B$, we observe that the anti-parallel states are shifted in $V_{\textrm{G1}}$ relative to the parallel states. For example, at $B = +40\,$mT, the total magnetic field $B_{\textrm{tot}}^{(1)}$ at QD 1 changes when the FSG magnetization state switches from $(+,+)$ to $(-,+)$. This consequently changes the transmission DoS of QD1 at $E_{\rm F}$, resulting in a shift of the bias triangle position in $V_{\textrm{G1}}$. We observe a similar shift in $V_{\textrm{G1}}$ for the magnetization states $(-,-)$ and $(+,-)$ at $B = -40\,$mT, thereby enabling us to optimize the TMR signal. 

\section{Analysis of a Third Resonance}

\begin{figure}[H]
	\centering
	\includegraphics[width= 1 \textwidth]{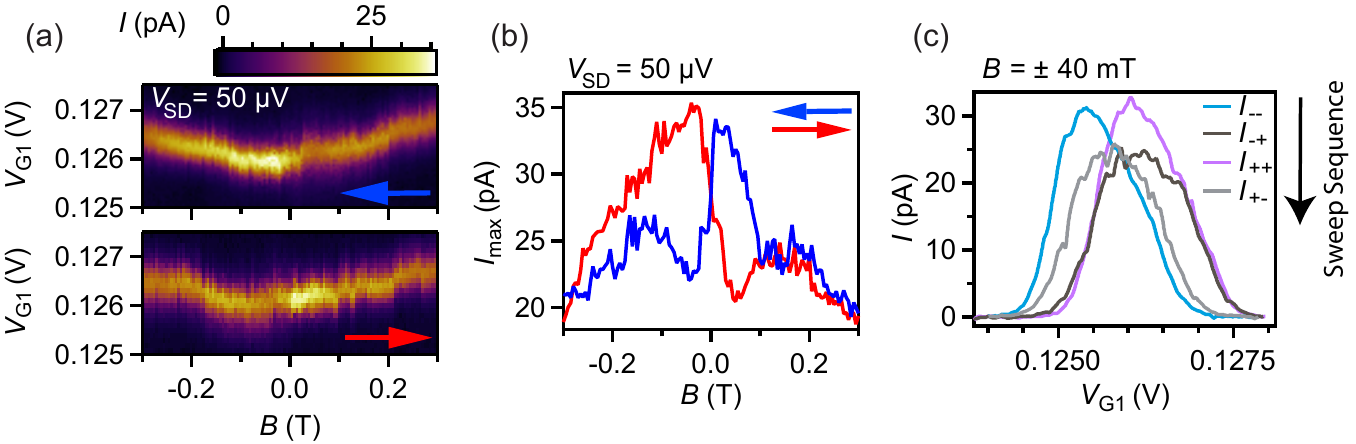}	
	\caption{\textbf{Data for resonance $\mathbf{R_{2}}$ in figure~1d of the main text.} (a) $I$ as a function of $V_{\textrm{G1}}$ and $B$ for the up (red) and down (blue) sweep for $V_{\textrm{SD}} = 50 \mu$V measured along the cross section $\textrm{C}_{3}$ in Figure 1d. (b) $I_{\textrm{max}}$ vs $B$ for the up and down sweeps extracted from \ref{add_data}a. (c) $I$ as a function of $V_{\textrm{G1}}$ for the four magnetization states measured at $B = \pm 40$ mT along the cross section $\textrm{C}_{4}$ in Figure 1d, parametrized by $V_{\rm G1}$.}
	\label{add_data}
\end{figure}

We present additional data (figure \ref{add_data}a) for the cross section $\textrm{C}_{3}$  of bias triangle $\textrm{R}_{2}$ in Figure~1d of the main text. $I$ as a function of $V_{\textrm{G1}}$ and $B$ are clearly hysteretic for the up and down sweep, mirrored around $B = 0$. The hysteresis is clearly visible in the $I_{\textrm{max}}$ vs $B$ curves (figure \ref{add_data}b) extracted from figure \ref{add_data}a, showing similar characteristics as the resonance in figure 3b. In addition, we measure $I$ as a function of $V_{\textrm{G1}}$ for each magnetization state at $B = \pm 40$ mT along the cross section $\textrm{C}_{4}$ (figure~1d), similar to figure~3c in the main text. The $(-,-)$ [blue] and $(+,-)$ [grey] magnetization states in Figure~\ref{add_data}c were measured at $B = -40$ mT, while the $(+,+)$ [purple] and $(-,+)$ [black] magnetization states were measured at $B = +40$ mT. Similar to Fig 3c, we observe a suppression of $11\%$ in $I$ for the anti-parallel magnetizations relative to the parallel ones.

\begin{figure}[H]
	\centering
	\includegraphics[width= 0.3 \textwidth]{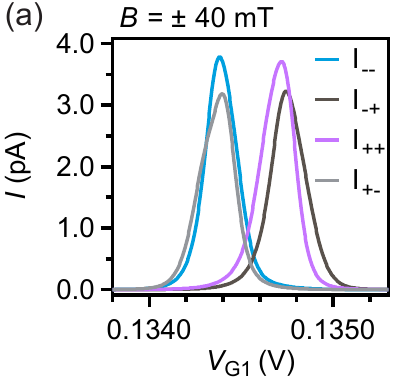}	
	\caption{\textbf{Resonant tunneling model.} (a) $I$ as a function of $V_{\textrm{G1}}$ extracted from the resonant tunneling model for all four magnetization states at $B = \pm 40\,$mT, reproducing experiments in Figure~3c of the main text. The $(-,-)$ [blue] and $(+,-)$ [grey] magnetization states are measured at $B = -40$ mT, while the $(+,+)$ [purple] and $(-,+)$ [black] magnetization states measured at $B = +40$ mT.}
	\label{model_data}
\end{figure}

The resonances $\textrm{R}_{1}$ and $\textrm{R}_{2}$ as well as resonance $\rm A$ in Figure~2 in the main text can be reproduced by the resonant tunneling model with very similar parameters, as summarized in table~S1. The $B_{\rm str1}$ and $B_{\rm str2}$ values mentioned in the main text are consistent with all the observed resonances. The model curves for the four magnetization states at $B = \pm 40\,$mT along cross section $\rm C_{2}$ (figure~1d in the main text), reproducing the experiments of Figure~3c is shown in Figure~\ref{model_data}.

\vspace{5 mm}

\begin{table}[h]
	\begin{center}
		\begin{tabular}{c||c|c|c|c|c|c|}
			Resonance & $g_{1}$ & $g_{2}$ &   $E_{2}^{(0)}$ ($\mu$eV) \\
			\hline
			$\textrm{R}_{1}$ Up sweep   & 5.2 - 5.9 & 6.1 - 6.5 & 7.5 - 8.5\\
			$\textrm{R}_{1}$ Down sweep & 5.0 - 5.6 & 6.1 - 6.5 & 7.5 - 8.5\\
			$\textrm{A}$ (Figure 2)		& 5.8 - 6.5 & 5.0 - 6.0 & 8.0 - 13.0 \\
			$\textrm{R}_{2}$ 			& 5.0 - 5.6 & 5.1 - 5.4 & 8.0 - 9.0 \\ 
		\end{tabular}
		\caption{Summary of the parameters extracted from the resonant tunneling model for the three resonances measured (g-factors and energy offset $E^{(0)}$ of the two QDs).}
	\end{center}
	\label{table}
\end{table}
